\documentclass[a4paper,fleqn,usenatbib]{mnras}
\usepackage[T1]{fontenc}
\usepackage{ae,aecompl}
\usepackage{graphicx}	
\usepackage{amsmath}	
\usepackage{amssymb}	
\usepackage{times}
\usepackage{epsfig}
\usepackage{natbib}
\usepackage{longtable}
\usepackage{float}
\usepackage{multirow}
\usepackage{xcolor}
\usepackage{hyperref}

\def\kpc{{\ h^{-1} \ \rm kpc}}
\def\kms{{\ \rm km \ s^{-1}}}

\def\apj {ApJ}
\def\apjl {ApJL}
\def\apjs {ApJS}
\def\aj {AJ}

\def\aap {A\&A}
\def\mnras {MNRAS}

\title[Satellites and central galaxies in SDSS]{Satellites and central  galaxies in SDSS: the influence of interactions on their properties} 

\author[Mesa et al.]{
Valeria Mesa,$^{1,2}$\thanks{E-mail: vmesa@mendoza-conicet.gob.ar} 
Sol Alonso,$^{3}$ 
Georgina Coldwell,$^{3}$ 
Diego Garc\'ia Lambas $^{4}$ 
\newauthor
and J.L. Nilo Castellon $^{1}$
\\
$^{1}$ Departamento de Astronom\'{i}a, Facultad de Ciencias, Universidad de La Serena, Av. Juan Cisternas 1200 Norte, La Serena, Chile\\
$^{2}$ Instituto Argentino de Nivolog\'{i}a Glaciolog\'{i}a y Ciencias Ambientales (IANIGLA-CONICET), Parque Gral San Mart\'{i}n, CC 330, CP 5500, Mendoza, Argentina\\
$^{3}$ 
Departamento de Astronom\'ia y Geof\'isica, CONICET, Facultad de Ciencias Exactas, F\'isicas y Naturales (FCEFN)--UNSJ, Av. Jos\'e Ignacio de la Roza Oeste 590, \\
J5402DCS, San Juan, Argentina  \\
$^{4}$
Instituto de Astronom\'{\i}a Te\'orica y Experimental, (IATE-CONICET),
Laprida 854, X5000BGR, C\'ordoba, Argentina}

\date{Accepted XXX. Received YYY; in original form ZZZ}

\pubyear{2020}

\begin{document}
\label{firstpage}
\pagerange{\pageref{firstpage}--\pageref{lastpage}}
\maketitle
\begin{abstract}
We use SDSS-DR14 to construct a sample of galaxy systems consisting of a central object and two satellites. 
We adopt projected distance and radial velocity difference criteria and impose an isolation criterion to avoid membership in larger structures. We also classify the interaction between the members of each system through a visual inspection of galaxy images, finding  $\sim80\%$ of the systems lack evidence of interactions whilst the remaining $\sim20\%$ involve some kind of interaction, as inferred from their observed distorted morphology. 
We have considered separately, samples of satellites and central galaxies, and each of these samples were tested against suitable control sets to analyse the results. 
We find that central galaxies showing signs of interactions present evidence of enhanced star formation activity and younger stellar populations. As a counterpart, satellite samples show these galaxies presenting older stellar populations with a lower star formation rate  than the control sample. The observed trends correlate with the stellar mass content of the galaxies and with the projected distance between the members involved in the interaction. The most massive systems are less affected since they show no star formation excess, possibly due to their more evolved stage and less gas available to form new stars. Our results suggest that it is arguable a transfer of material during interactions, with satellites acting as donors to the central galaxy. As a consequence of the interactions, satellite stellar population ages rapidly and new bursts of star formation may frequently occur in the central galaxy.
\end{abstract}


\section{Introduction}

Throughout the history of the Universe, galaxy-galaxy interactions play a crucial role in galaxy formation and evolution according to the hierarchical model of structure formation, by linking together star formation processes with galaxies growth \citep{Alonso2006, Woods2007, Ellison2010, Lambas2003, Lambas2012, Mesa2014}. 
On the observational side, several works \citep[e.g.][]{Yee1995, Kennicutt1998, Rogers2009, Ellison2011, Lambas2012} have shown that interactions between galaxies are powerful mechanisms to trigger \textit{star formation activity (SFA)}. Taking into account statistical analysis of galaxy pairs, \cite{Barton2000} and \cite{Lambas2003} have found that the proximity in radial velocity and projected distance is correlated to an increase of the SFA. Also, \cite{Balogh2004} found a correlation between star formation rate (SFR) and environment attributable to starbursts induced by galaxy-galaxy interactions.

In this scenario, the interaction between galaxies of similar size and mass would be a very good laboratory to study the consequences of the fusion processes in the formation of galaxies. Unfortunately, these systems are statistically insignificant, focusing the studies in systems with higher mass, accompanied by the so-called satellite galaxies.  While galaxy collisions are expected to be more violent, encounters between galaxies with their smaller companions, would be the most common, because low-light galaxies are more frequent in the Universe. In this line, \cite{Daddi2005} using HST data to study the evolution of early type galaxies, found signatures in the B band, compatible with an ongoing merger or cannibalism of satellites. 
\cite{Trujillo2006,Trujillo2007} have studied the size evolution of compact massive galaxies, finding that dry merger scenario can be considered as a reasonable mechanism for the subsequent evolution of these galaxies, since this type of mergers are not efficient at forming new stars, but are efficient in increasing the size of the objects.  These results have also been reported by \cite{vanDokkum2008}  using deep and high-resolution  images and moreover by \cite{vanderWel2014} through the use of spectroscopy  and  multiwavelength photometry from the 3D-HST survey combined with CANDELS imaging. In addition, \cite{Bernardi2009} showed that early-type BCGs identified in the Sloan Digital Sky Survey (SDSS) grew from many dry minor mergers. More recently, \cite{Vulcani2014} analysed the relation between colour and structure within galaxies using GAMA survey, showing that early type galaxies are associated with multiple collapse and merging events. 

There are different and varied studies about the impact of the fusion of galaxies in these systems. They show that the presence of a close companion generates a clear increase in the morphological asymmetries of the galaxy even at 50 $ \kpc $ away \citep{ Patton2016}. Numerical simulations have shown that galaxies grows by accreting other galaxies, mostly smaller companions \citep{Shao2018, Nipoti2018, Forbes2016}. \cite{Naab2009} have studied the influence of minor mergers on the evolution of elliptical galaxies. Their results show that this type of encounters would be the drivers for the late evolution of sizes and densities of early-type galaxies. In this sense, \cite{Oser2010} also provides that the formation of galaxies has different phases, and that an extended phase in evolution consists of an important growth due to the accretion of smaller satellite stellar systems. On the other hand, \cite{Hirschmann2015} showed that stellar accretion from minor mergers of satellite galaxies results in steep negative metallicity and colour gradients and slightly positive age gradients, successfully matching the observed profiles of local galaxies.
All this allows us to conclude the importance of the study of satellite galaxy systems and the analysis of their properties.

\cite{Stierwalt2015} presented a systematic study about star formation and the further processing of the interstellar medium in the interaction between dwarf galaxies.
The authors concluded that the interactions between dwarf  galaxies are important conductors of galactic evolution in the low mass end, but ultimately the environment is responsible for the extinction of star formation.

For all environments, bulge-dominated galaxies have a colour-magnitude diagram dominated by red galaxies which depends linearly on absolute magnitude \citep{Hogg2004, Blanton2005}. 
Many authors have also studied the relationship of satellite galaxies with their environment, based on numerical simulations. For instance, \cite{Barber2015} predict a statistical excess of satellites whose main axis aligns with the direction to the central galaxy. Evidence of this was found in the  satellite population of M31, which suggests that tidal effects may have played an important role in its evolution. \cite{Tempel2015} also noted this alignment, suggesting that filaments identified on larger scales can be reflected in the positions of the satellite galaxies that are very close to its central galaxies. \cite{Sales2015} examined the colours of satellite galaxies in the Illustris simulation. They found that the satellites roughly trace the distribution of dark matter in their system, and that in massive systems, red satellites dominate and are distributed more abruptly than the blue population, while for the lower mass primary galaxies, the satellites are mostly blue.
Additionally, observational studies suggest that the properties of a satellite galaxy are strongly correlated with those of its central galaxy  \citep{Weinmann2006}. Using a large galaxy group catalogue constructed from the SDSS, \cite{vandenBosch2008} have proved that satellites are redder and slightly more concentrated than central galaxies of the same stellar mass. This scenario points to strangulation as the main mechanism that operates on satellite galaxies, and that causes their transition from the blue to the red sequence. These results were also reported by \cite{Wetzel2012} from SDSS data too. 

Furthermore, \cite{Deason2014} research the frequency of major mergers between dwarf galaxies in the Local Group using cosmological simulations. They found that $\sim10\%$ of dwarf satellite galaxies  with  $M*> 10^{6}M\odot$ inside the virial radius, experienced a major fusion with a ratio of stellar mass close to 0.1 from $z=1$, with a lower fraction for dwarf galaxies of smaller mass. They found that satellite-satellite mergers also occur within the main halo after virial infall, catalysed by the large fraction of dwarf galaxies that fall down into the group. The fraction of fusions doubles for dwarf galaxies outside the virial radius as well  that the most distant dwarf galaxies in the local group are the most likely to have experienced a recent major merger.
\cite{Tinker2010} tested how galaxies evolve onto the red sequence, finding that $\sim$ 60$\%$ of satellite galaxies being red or quenched, involving that $\sim$ 1/3 of the red sequence is comprised of satellites.  

\cite{Wetzel2013} examined the star formation histories of the satellite galaxies from SDSS data, together with a high-resolution N-body simulation, finding a delayed-then-rapid quenching scenario. In the similar direction, \cite{Hirschmann2014} showed that satellite galaxies with internal suppression of star formation activity, could be experimenting AGN/radio-mode/stellar feedback. \cite{Oman2016} found that quenching of satellite galaxies by massive clusters is 100$\%$ efficient, and also showed that all satellites quench on 
their first infall.

More recently, \cite{Delucia2019} used semi-analytic models to study the time-scales in which star formation is suppressed in satellite galaxies. Finding that environmental processes play a marginal role in passive galaxies with stellar mass larger than $M*> 10^{10}M\odot$. However, the models need to be improved to predict the behaviour of less massive galaxies, as is the case of satellite galaxies.

There are few observational studies on this topic, and in those found in the literature, the relation between satellites and their central galaxy stands out, but without a deep analysis on the possible interaction between satellite galaxies and how these influences the global system properties.
\cite{Gutierrez2006} studied a sample of 31 satellites orbiting isolated giant spiral galaxies finding three cases of clear interactions between the satellites.
Four of the galaxies in their sample are among the objects with higher starforming activity.
In contrast, the only two galaxies of the sample that are not forming stars are also members of these pairs. They propose that the presence of the bridges connecting a satellite with their companions, and the comparatively large amount of gas are signs of mass transfer from one galaxy to the other. This is probably inhibiting the star formation activity in the donor and enhancing it in the accreting galaxy. The stripping suffered by the satellite galaxies could also be responsible for its morphological changes.  Also, \cite{Knobel2013} showed that  the fraction of satellite galaxies that are red, is systematically higher than that of centrals, and that the satellite quenching efficiency (i.e. the probability that a satellite is quenched because it is a satellite rather than a central) is independent of stellar mass. These effects are likely to remain even at high redshifts.
More recently, \cite{Pasquali2019} examined the physical properties of satellite galaxies in the projected phase-space of their host environment, for satellites inside one virial radius of their host. They show that low mass satellites are more sensitive to environment and that the general characteristics depend on the time spent in their host environment. 

This paper is structured as follows: Section 2 describes the data used in this work. In Section 3 we show the procedure used to construct the satellite galaxies catalogue, explaining the classification process of the different types of interactions, and we also present the procedure for building the control samples.
An analysis of star formation rates, colours and stellar population, and their differences with the control samples is described in Section 4. Finally in Section 5, we summarise the main conclusions. 

Throughout this paper we adopt a cosmological model characterised by the parameters $\Omega_m=0.3$, $\Omega_{\Lambda}=0.7$ and $H_0=70 \kms \rm Mpc ^{-1}$.

\section{Data}

This work is based on data provided by the the Sloan Digital Sky Survey \citep[SDSS;][]{York}, one of the most successful surveys in the history of astronomy. Over years of operations (SDSS-I, 2000-2005; SDSS-II, 2005-2008; SDSS-III, 2008-2014) SDSS data have been annually released to the scientific community. The latest generation of the SDSS data \citep[SDSS-IV, 2014-2020;][]{sdssiv} is extending precision cosmological measurements to a critical early phase of cosmic history (eBOSS), expanding its
galactic spectroscopic survey to the north and south hemispheres (APOGEE-2), using for first time the Sloan spectrograph, performing spatially resolved maps of individual galaxies (MaNGA). In the present work we consider spectroscopic data from SDSS Data Release 14 \citep[][SDSS-DR14]{dr14}. For this sample, k-corrections band-shifted to $z=0.1$ were calculated using the software \texttt{k-correct\_v4.2} of \cite{Blanton2007}. For the data set, k-corrected absolute magnitudes were calculated from Petrosian apparent magnitudes converted to the AB system. In our analysis we will use the u, g and r-bands in the \textit{ugriz} system.

 In this paper we will carry out an analysis about the star formation efficiency, colour distributions, age of stellar populations based on different parameters such as the $D_n(4000)$ spectral index, $SFR/M*$, all available in the SDSS spectroscopic database. We obtain  all data  catalogues  through  SQL  queries  in  CasJobs\footnote{ http://skyserver.sdss.org/casjobs/}. From this catalogue we use the star formation rate normalised to the total mass in stars estimated from the SDSS fibre, $log(SFR/M_*)$, taken from \cite{Brinchmann2004}. As it has been discussed by these authors, aperture effects could be important for the most massive galaxies. Therefore, for satellite galaxies it is not expected to be an issue here, but this effect must be taken it into account for centrals. 
 We have compared the angular size of the SDSS fibre and the radius containing half of the galaxy light, ($r_{50}$). We find that for the majority of our centrals, the fiber size is within a 50$\%$ fraction of this radius. For this reason, we have compared the derived fibre SFR with the values obtained for the global SFR, as calculated by \cite{Brinchmann2004} finding no significant differences in the resulting values. So, although for central galaxies the fibre SFR estimate corresponds only to a small central portion, it provides reliable SFR global determinations. We use the total stellar masses $Log(M*/M\odot)$ calculated using the Bayesian methodology, and model grids described in \cite{Kauffmann2003}.
We also use the spectral index $D_n(4000)$, as an indicator of the age of stellar populations. 
This spectral discontinuity occurring at 4000\AA \ \citep{Kauffmann2003} arises by an
accumulation of a large number of spectral lines in a narrow region of the spectrum, an effect that is important in the spectra of old stars.
We have adopted \citet{Balogh} definition of  $D_n(4000)$  as the ratio of the average flux densities in the narrow continuum bands (3850-3950 \AA \  and 4000-4100 \AA). 
 Finally, to discriminate between bulge and disc-types galaxies, we use the concentration index C\footnote{$C=r90/r50$ is the ratio of Petrosian 90 \%- 50\% r-band light radii} \citep{Abraham1994}, a well tested morphological classification parameter  \citep{Strateva_2001},  also used as a good stellar-mass tracer ($M_*$) and an indirect index of the SFR \citep{Deng_2013}.  \citet{Yamauchi_2005} performed a galaxy morphological classification using the C parameter, finding a very good agreement with the visual classification.

\section{The Sample}

Our sample was created based on a combination of photometric and dynamical criteria.  First, a base sample composed of a bright galaxy plus fainter surrounding sources were defined. Galaxies brighter than  \textbf{ $M_r=-20.5 mag.$} were selected as central galaxies and objects  that lies inside $r_p<150 \kpc$  and $\Delta V<500 \kms$, restricted to a difference of 1.5 magnitudes fainter regarding to the central galaxy,  were classified as satellites of the main object. 
These criteria were chosen taking into account our previous experience in the study of galaxy pairs. For instance, \cite{Lambas2003,Alonso2006} found that  $r_p<100 \kpc$  and $\Delta V<350 \kms$ were convenient thresholds for stellar formation activity induced by the interactions, and  and that this is triggered in values lower than  $r_p<25 \kpc$  and $\Delta V<100 \kms$. On this basis, we then work with larger samples reaching up to  $r_p<50 \kpc$  and $\Delta V<500 \kms$ \cite{Mesa2014}, and $r_p<100 \kpc$  and $\Delta V<500 \kms$ \cite{Mesa2018}. Considering now the presence of another galaxy in the system, we decided to increase the projected distance $r_p$, taking into account the criteria used by \cite{Duplancic2018}  who used the value of $r_p<200 \kpc$ and $\Delta V<500 \kms$  to identify small galaxy systems.\\

At this stage, we only selected those systems with two satellites. In addition, was necessary to define isolation criteria in order to ensure that the dynamics of our systems is not dominated by larger virialised structures where they could be immersed. For instance, an adequate isolation criteria was taken into account considering that within a radius of 500 $\kpc$ and $\Delta V<1000 \kms$ there should not be a brighter galaxy than one magnitude fainter than the central galaxy. \cite{omill2012} adopted similar thresholds to find galaxy triplets and also \cite{Duplancic2018}, with the aim to define an homogeneous selection criteria of small galaxy systems.  Furthermore, the systems were requested to be within z$<$0.1. \\

\begin{figure}
  \centering
  \includegraphics[width=.4\textwidth]{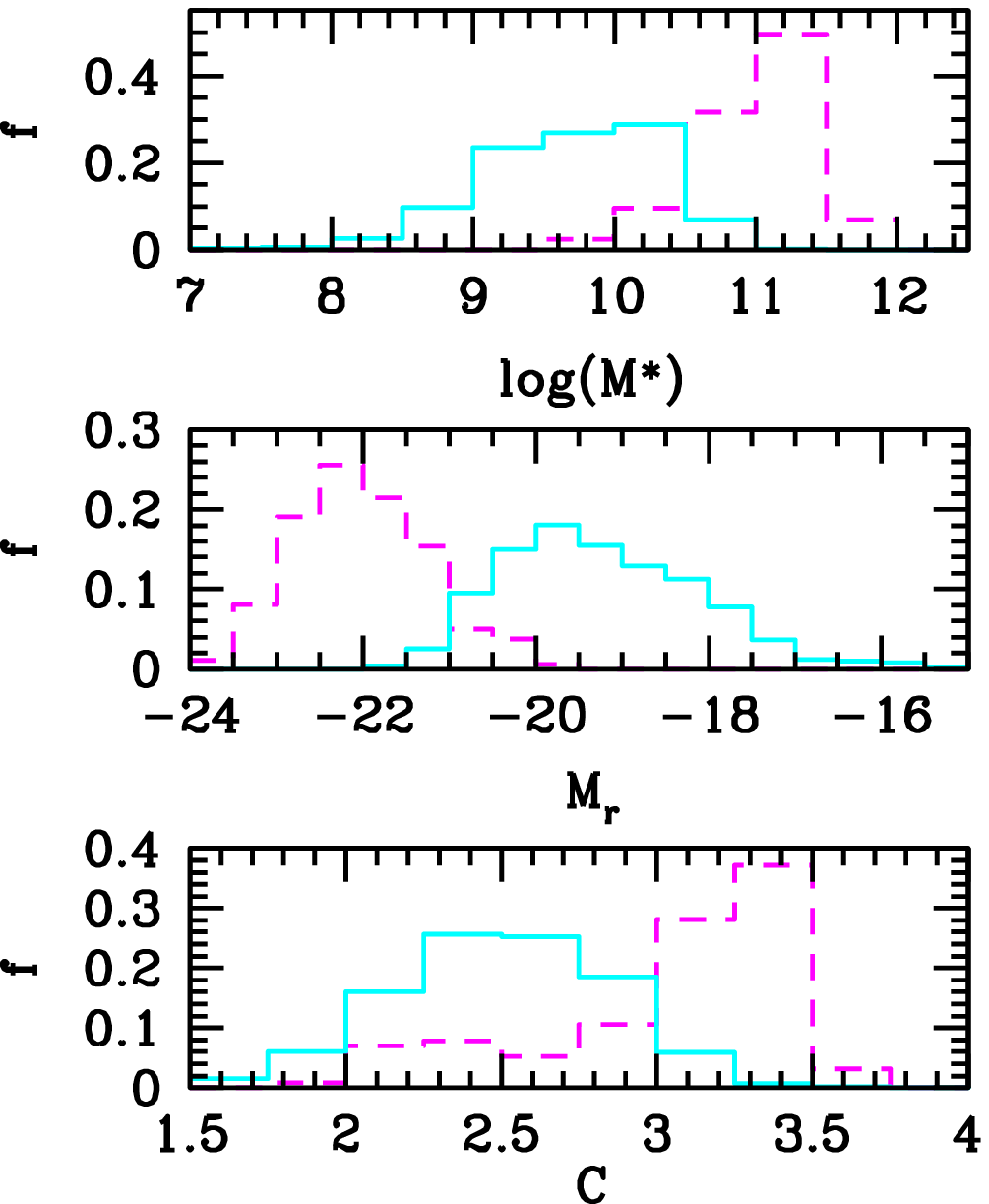}
  ~\hfill
  \includegraphics[width=.4\textwidth]{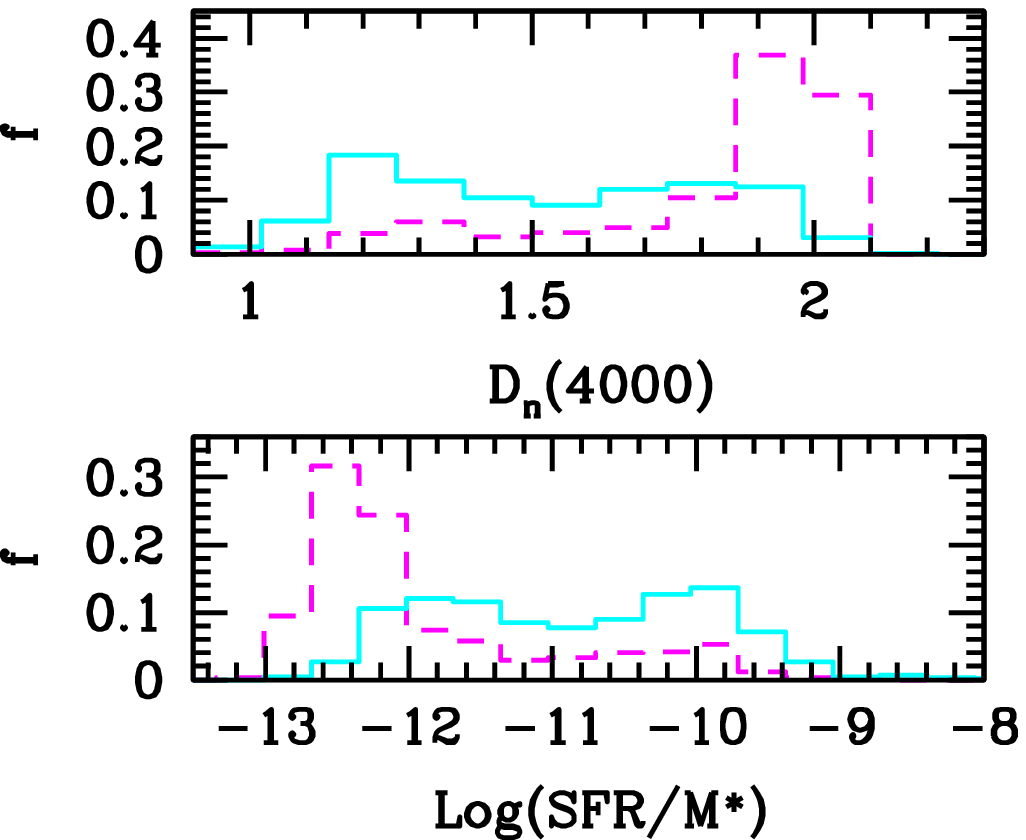}
  \caption{ From top to bottom: Distribution of $Log(M*)$, $M_r$, concentration index ($C$), $D_n(4000)$ and $Log(SFR/M*)$   for central galaxies (dashed lines) and satellites (solid lines).  }
  \label{mstar}
\end{figure}

\begin{figure}
  \centering
   \includegraphics[width=.45\textwidth]{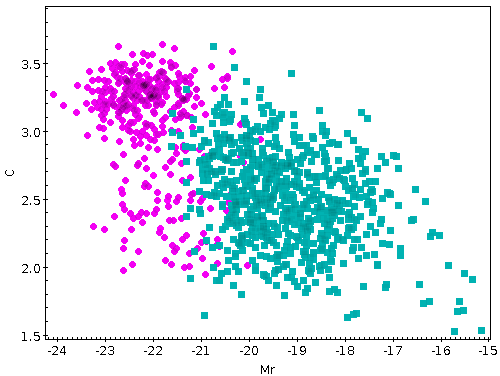}
  \caption{Concentration index ($C$) vs $M_r$ for central galaxies (magenta circles) and satellites (cyan squares). }
  \label{CMr}
\end{figure}

In Fig \ref{mstar} we show the normalised distributions of $Log(M*)$, r-band $M_r$, concentration index ($C$), $D_n(4000)$ and $Log(SFR/M*)$ for central galaxies and their satellites. In order to quantify these differences we have computed the mean values of  $Log(M*)$ finding  9.68$\pm$ 0.02 and 11.00 $\pm$ 0.01 for satellites and central galaxies respectively. In an analogous procedure for $M_r$ we find values of -19.23$\pm$ 0.04 and -22.02 $\pm$ 0.02. Is observed a difference of three magnitudes between main galaxies and satellites, probably related to the nature of the sample. In the case of concentration index ($C$) we found values of 2.51$\pm$ 0.01 and 3.04$\pm$ 0.01, for $D_n(4000)$ values corresponding to 1.51$\pm$ 0.01 and 1.82$\pm$ 0.01 and finally, for  $Log(SFR/M*)$ values of -10.85$\pm$ 0.03 and -11.89$\pm$ 0.03 respectively. From this analysis we can infer that we are faced with two populations with notable and obvious differences, which reveal their dependence on the morphology of the galaxies that compose our systems. Due to the conditions imposed to obtain the sample, we find very bright and massive central galaxies, and also more evolved ones. On the other hand, satellite showing more typical properties of late-type galaxies (keeping in mind that low values of $D_n(4000)$ and higher ones of $Log(SFR/M*)$ indicates active star formation activity and younger stellar populations).
Also, in Fig \ref{CMr} we show  concentration index ($C$) vs $M_r$ for central galaxies and satellites highlighting again the great differences between both samples. Different authors  \citep[e.g.][]{Gadotti2009,Mesa2014,Morselli2017} propose that an adequate threshold to separate galaxies between bulge-types and disc-type is 2.5. From this plot we can see that the satellite sample presents all kinds of morphology, conversely the central galaxies exhibit higher values of C index, indicating that a higher fraction of galaxies in this sample presents bulge morphology.  \\

\subsection{Classification}

Once the catalogue of satellite galaxies was obtained, we performed an eye-ball classification using the SDSS-DR14 imaging available in SkyServer\footnote{https://skyserver.sdss.org/dr14/en/tools/chart/listinfo.aspx} in order to distinguish between different classes of interactions. The systems were classified according to three categories:\\
(i) the main object with two satellites without apparent interactions, \\
(ii) mutual interaction between the satellites, or \\
(iii) between the main object and some of its satellites. \\

\begin{figure*}
	\centering
	\includegraphics[width=1\textwidth]{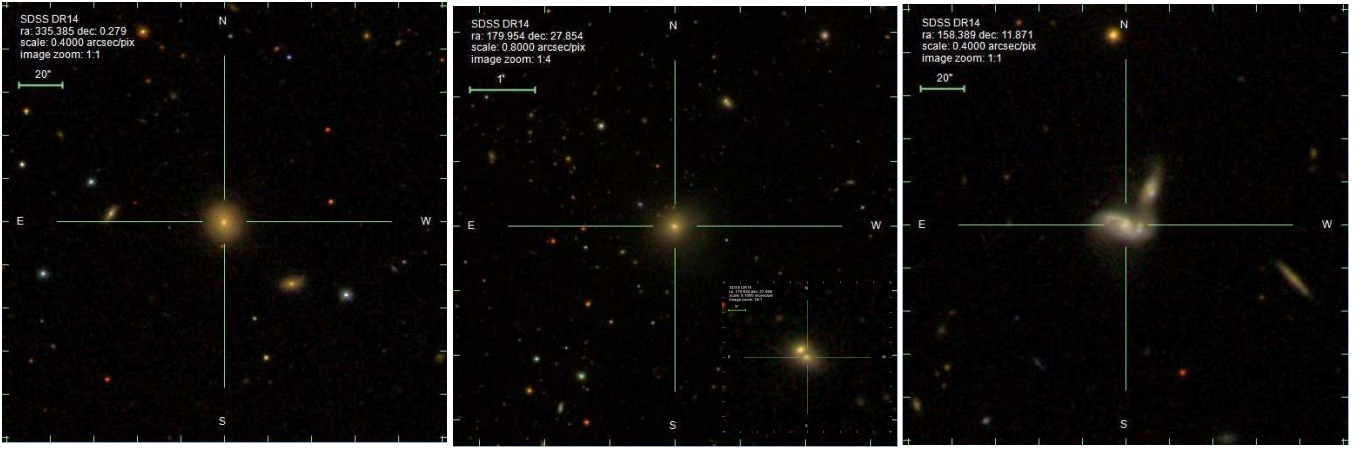}
	\caption{Examples of galaxy systems images with different classification: 
	Systems without interaction (left),interaction between satellites (middle) and interaction with main galaxy (right). The inbox in middle panel shows a zoom to the interacting satellites.}
	\label{fig:example}
\end{figure*}

This procedure was made with the purpose of analysing the relation between the different components of the systems. Fig. \ref{fig:example} shows examples of the different classifications. It is important to note on how different types of interactions influence on the accretion process of the material in primary galaxies. This visual classification is important since it  allows to classify different type of interactions and besides it permits  to detect spurious systems and/or misclassification from SDSS. We have used this method previously with excellent results \citep[e.g.][]{alonso2007, Lambas2012, Mesa2014}. This technique also allows us to clean the sample of galaxies immersed in groups, undetected by the software due the galaxies only have photometric information. We found that 94\% of the systems were classified into these subsamples. The remaining that do not fulfil these three categories were excluded from the present analysis. Table \ref{t1} provides the classification, number of systems and percentages in this sample of satellite galaxies.

\begin{table}
\center
\caption{Classification, number of systems and percentages in the different types.}
\begin{tabular}{| l c c | }
\hline
Classification & Number of systems & Percentages\\
\hline
Non Interacting      &    338    & 80.67\%  \\
Interaction between satellites      &  3      &   0.72 \% \\
Interaction satellites central galaxy     &       78 &    18.62\% \\             
Total   &    419      & 100\% \\

\hline
\end{tabular}
	\label{t1}
\end{table}

\subsection{ Control samples}
In order to understand the behaviour of our systems, we used a control sample for each catalogue, central and satellite galaxies, with the aim to compare different properties with respect to isolated galaxies. Therefore we use a Monte Carlo algorithm we build control samples of galaxies without a companion by matching the redshift ($z$), r-band absolute magnitude ($M_r$) and concentration index ($C$) distributions of our samples, following the work of \citep{Perez_2009}. The process was done simultaneously for each parameter, randomly matching bins of 0.5 mags for Mr, 0.015 for z and 0.25 for C, respectively. These control samples have a larger number of galaxies than the main samples allowing to have confident statistical testing sets.\\

\begin{figure*}
  \centering
  \includegraphics[width=.5\textwidth]{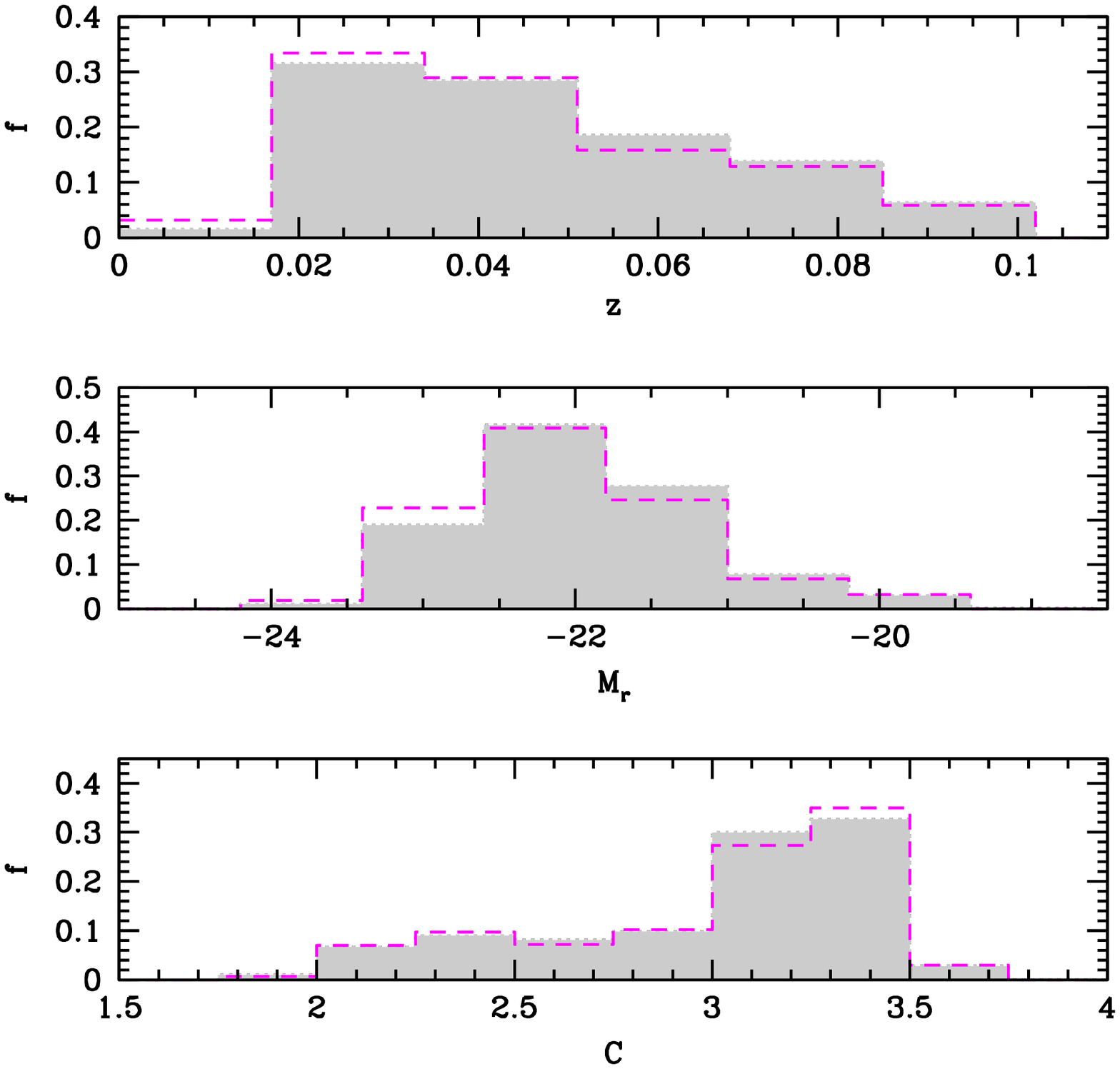}~\hfill
  \includegraphics[width=.5\textwidth]{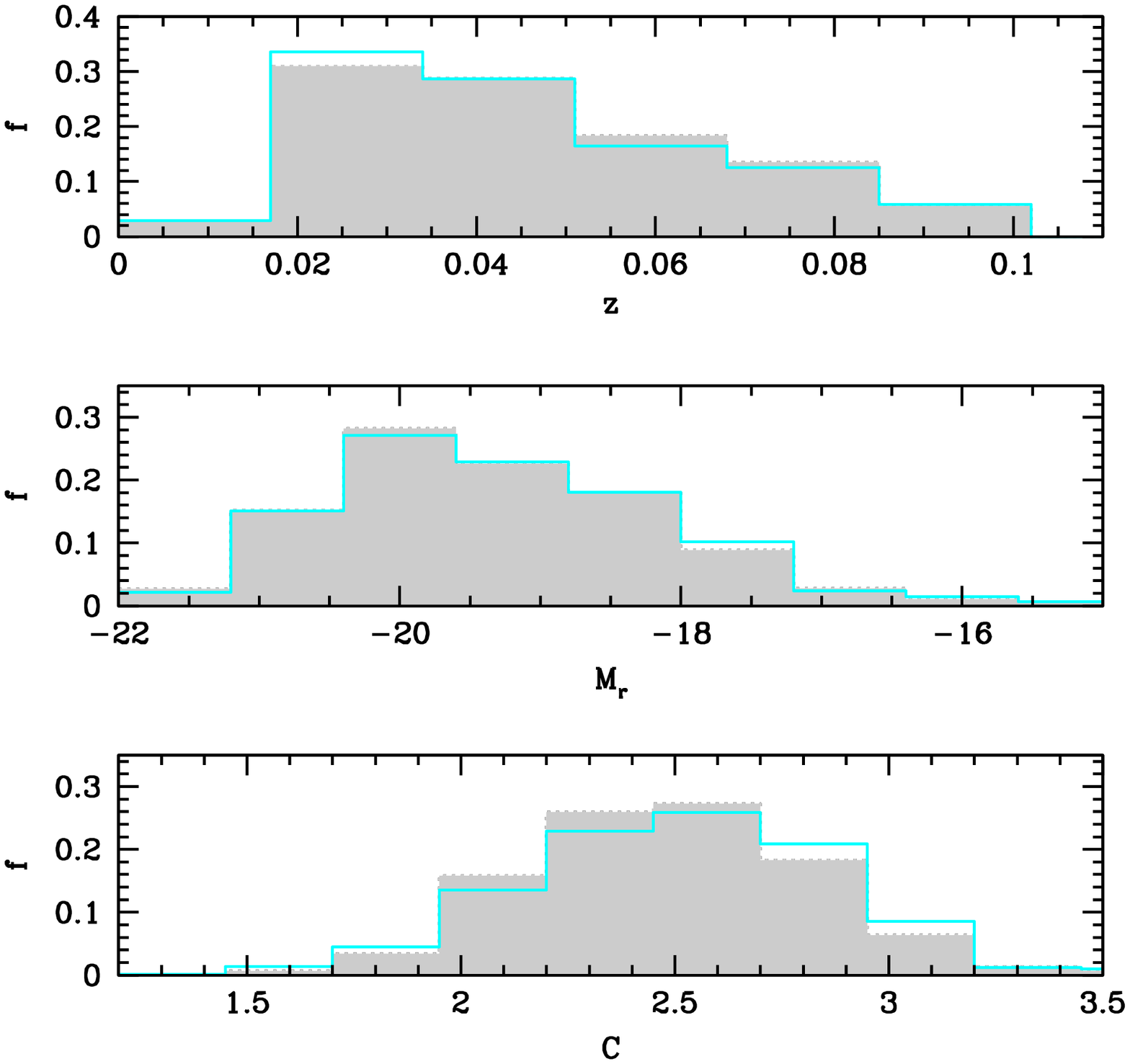}
  \caption{{Left:}Distribution of z, $M_r$ concentration index ($C$) for central galaxies (dashed lines) and control sample (shaded). \protect\\{ Right:} Satellites (solid lines) and control sample (shaded).}
  \label{prop_c}
\end{figure*}

We build control samples matching  the values of the parameters listed above (Fig. \ref{prop_c}). This procedure was performed separately, both for central galaxies and their satellites. 
In all cases we performed a Kolmogorov-Smirnov (KS) test and we obtained $p>0.05$, hence we can not reject the null hypothesis that the samples were drawn from the same distribution. Then, any difference in the galaxy properties is associated only with the interaction, consequently, by comparing the results we estimate the real difference between satellite or central galaxies (with possible interactions)  and galaxies without a companion, unveiling the effect of morphology or luminosity on this features.
\\

\section{Analysis}
\subsection{Central galaxies}

Through this work, it is intended to have a greater knowledge of how the properties of the primary galaxies are affected by the presence of their satellites, as well as the role of the interactions on them.

To this, we will focus on the analysis of the properties of the central galaxies of our systems under study. For this purpose, the systems have been split into two groups, those with the interaction between the satellites together with those with signatures of interaction between the central galaxy and the satellite, with the aim to have better statistics. And on the other hand, we have considered those systems without obvious interaction.

\subsubsection{Colours}

\begin{figure*}
  \centering
  \includegraphics[width=.50\textwidth]{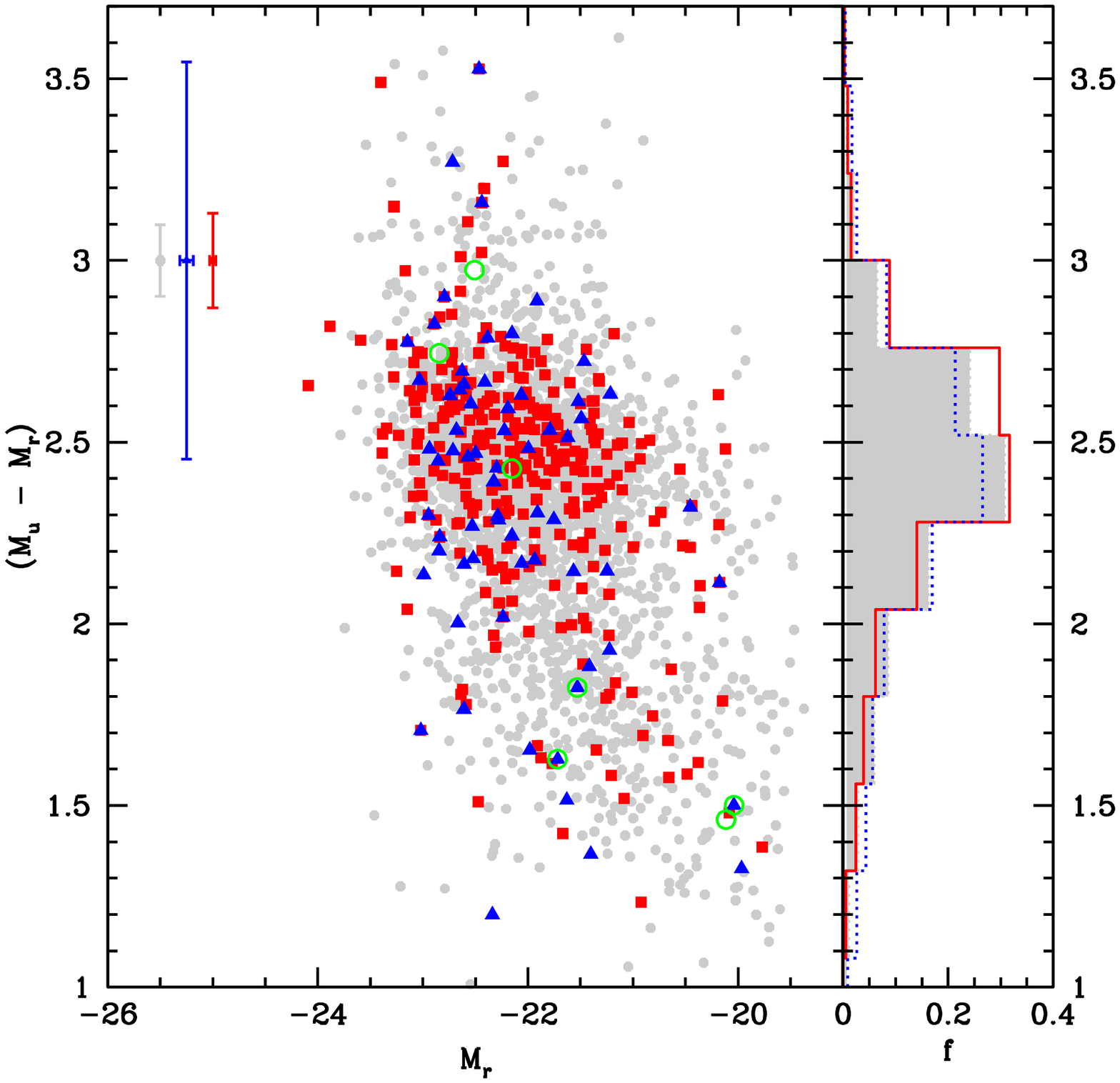}~\hfill
  \includegraphics[width=.50\textwidth]{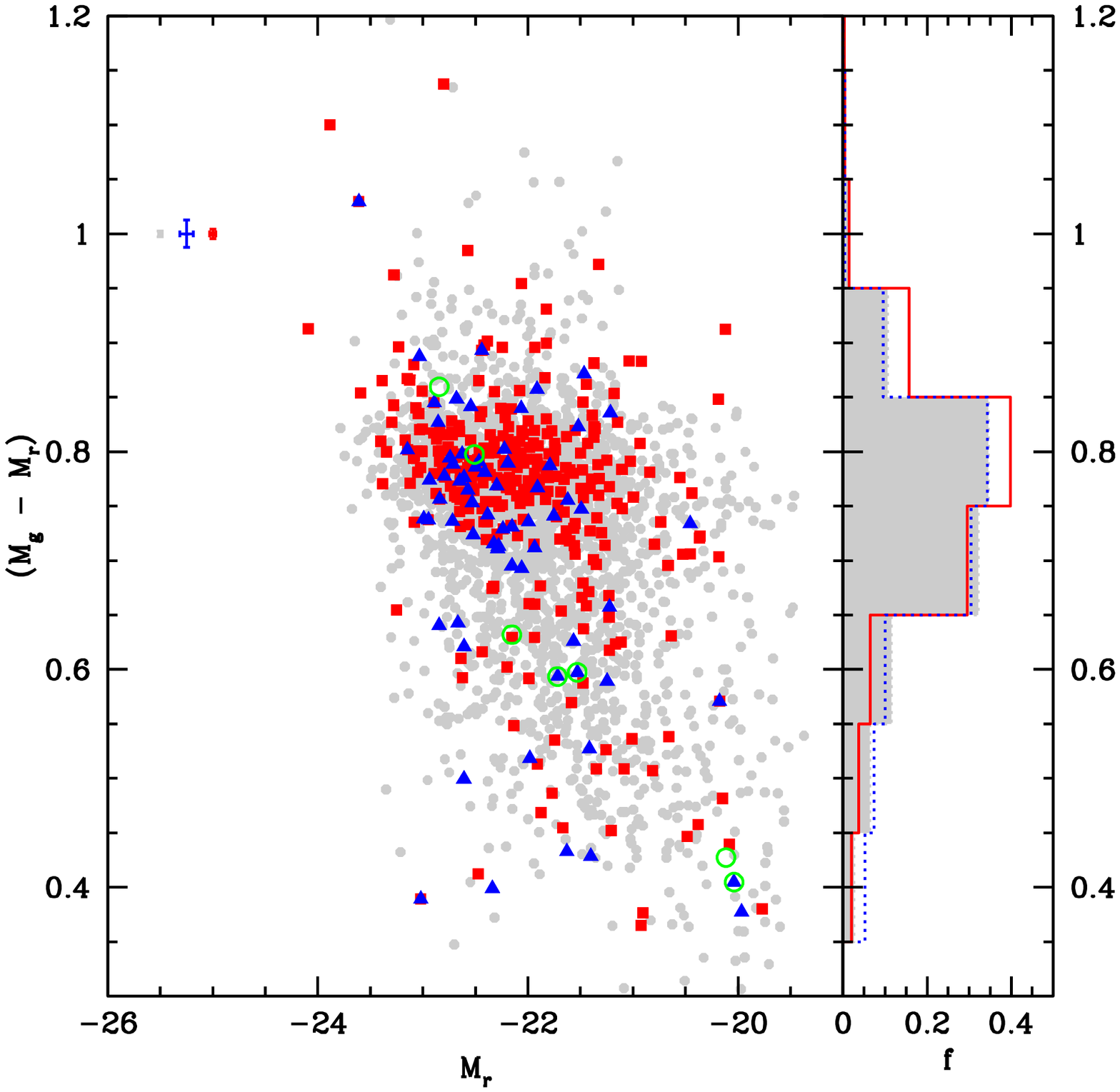}
 \caption{{Left:} Colour magnitude diagram for central galaxies in systems without interaction (red squares) and systems with interaction between satellites or with main galaxy (blue  triangles) and control sample (grey dots), and $Mu-Mr$ normalised distribution. Green open circles represents central galaxies in systems with double interactions. Points in the upper left corner represent data errors, for each subsample. \protect\\{ Right:} Colour magnitude diagram  and $Mg-Mr$ normalised distribution.}
 
   \label{colM}
\end{figure*}

In order to characterise the colours of the central galaxies belonging to our sample, in figure \ref {colM} the colour-magnitude diagrams ($ M_u-M_r $ and $ M_g-M_r $ versus $M_r$) of these galaxies are observed, with and without interaction with their satellites. The comparison sample is also included. 
We can observe that central galaxies are mostly populating the so-called "red sequence", while the galaxies with interaction and those belonging to the control sample, tend to be located in the region of the "green valley". This behaviour is also reflected in the colour index distributions, $ M_u-M_r $ and $ M_g-M_r $, in the same figure.

\begin{figure}
 \centering
 \includegraphics[width=.4\textwidth]{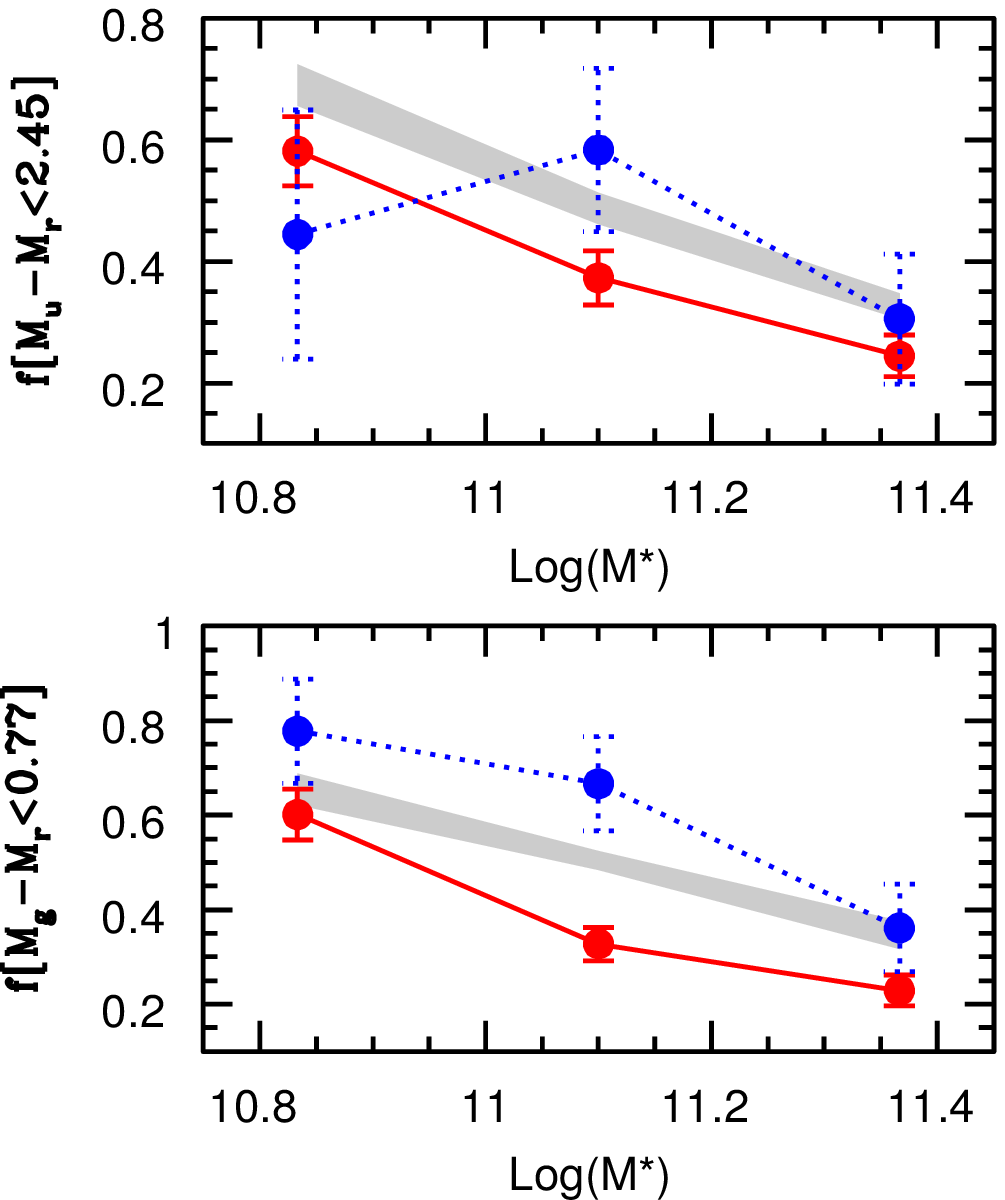}
\caption{Fraction of galaxies with $M_u-M_r$ and $M_g-M_r$  smaller than the median of the sample for central galaxies in systems without interaction (red solid) and systems with interaction between satellites or with main galaxy (blue dotted) and control sample (shaded). }
  \label{frac_ur}
\end{figure}

The fraction of objects with bluer colours than the median of the sample has been calculated. Fig \ref{frac_ur} shows these fractions as a function of $Log(M*)$, exhibiting that the central galaxies with interactions have a slightly higher fraction of blue colours with respect to the other samples. All the uncertainties  were derived through a bootstrap resampling technique \citep{Barrow}.

\subsubsection{Star formation and stellar populations}

\begin{figure}
 \centering
\includegraphics[width=.50\textwidth]{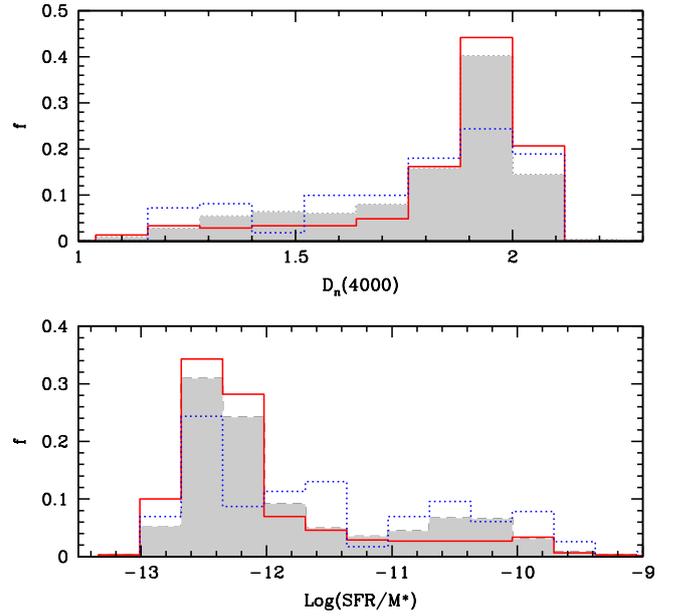}
\caption{Distribution of $D_n(4000) $  spectral index and $Log(SFR/M_*)$ for central galaxies in systems without interaction (red solid) and systems with interaction between satellites or with main galaxy (blue dotted) and control sample (shaded).}
  \label{dn_c}
\end{figure}

To study the age of the stellar populations, the spectral index, $D_n (4000)$, and the specific star formation rate, $Log(SFR/M*)$, will be used. The standard distributions of these parameters for each type of system are shown in Fig \ref{dn_c}, together with their control sample defined in previous Section. It can be seen that the central galaxies of our systems constitute an old population aged and with low star formation. 
With some differences only in the case to have signatures of interactions, the central galaxies of the systems show more efficient star formation activity and younger stellar population, with respect to the control samples.
 
To quantify these differences, we have calculated the fraction of young galaxies with stellar formation, that is, the fraction of objects with values of $D_n (4000)$ that are below the median of the total sample, and values of $Log(SFR/M*)$ above. This also allows to rule out the influence that the masses of galaxies may have.

\begin{figure}
 \centering
 \includegraphics[width=.4\textwidth]{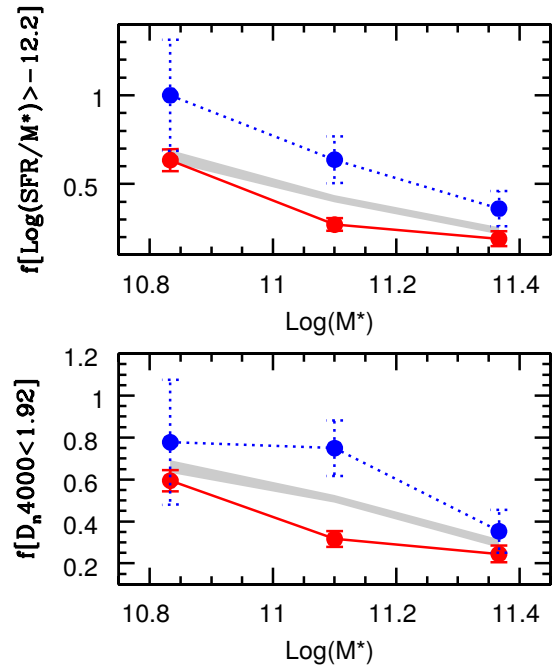}
\caption{Fraction of galaxies with $Log(SFR/M_*)$ higher and $D_n(4000) $ lower than the median of the sample for central galaxies in systems without interaction (red solid) and systems with interaction between satellites or with main galaxy (blue dotted) and control sample (shaded).}
  \label{frac1}
\end{figure}

In Fig \ref{frac1} these fractions are shown as a function of the mass of the galaxies. In this case, there are notable differences in the behaviour of galaxies with or without interactions with their satellites, the former being the ones with the highest fractions, decreasing as the mass of the objects increases. The control sample shows an intermediate behaviour between these two. At greater mass, no differences are observed between the samples, within the errors considered.

\subsection{Satellite galaxies}

This section is devoted to the study of the properties of the satellite galaxies in our sample, in order to understand how and to what extent they are affected by the processes of interaction between them or with their central galaxy. Similarly to the previous section, they have been grouped into two categories: those with some kind of interaction and those with no apparent interaction.

\subsubsection{Colours}
\begin{figure*}
  \centering
  \includegraphics[width=.50\textwidth]{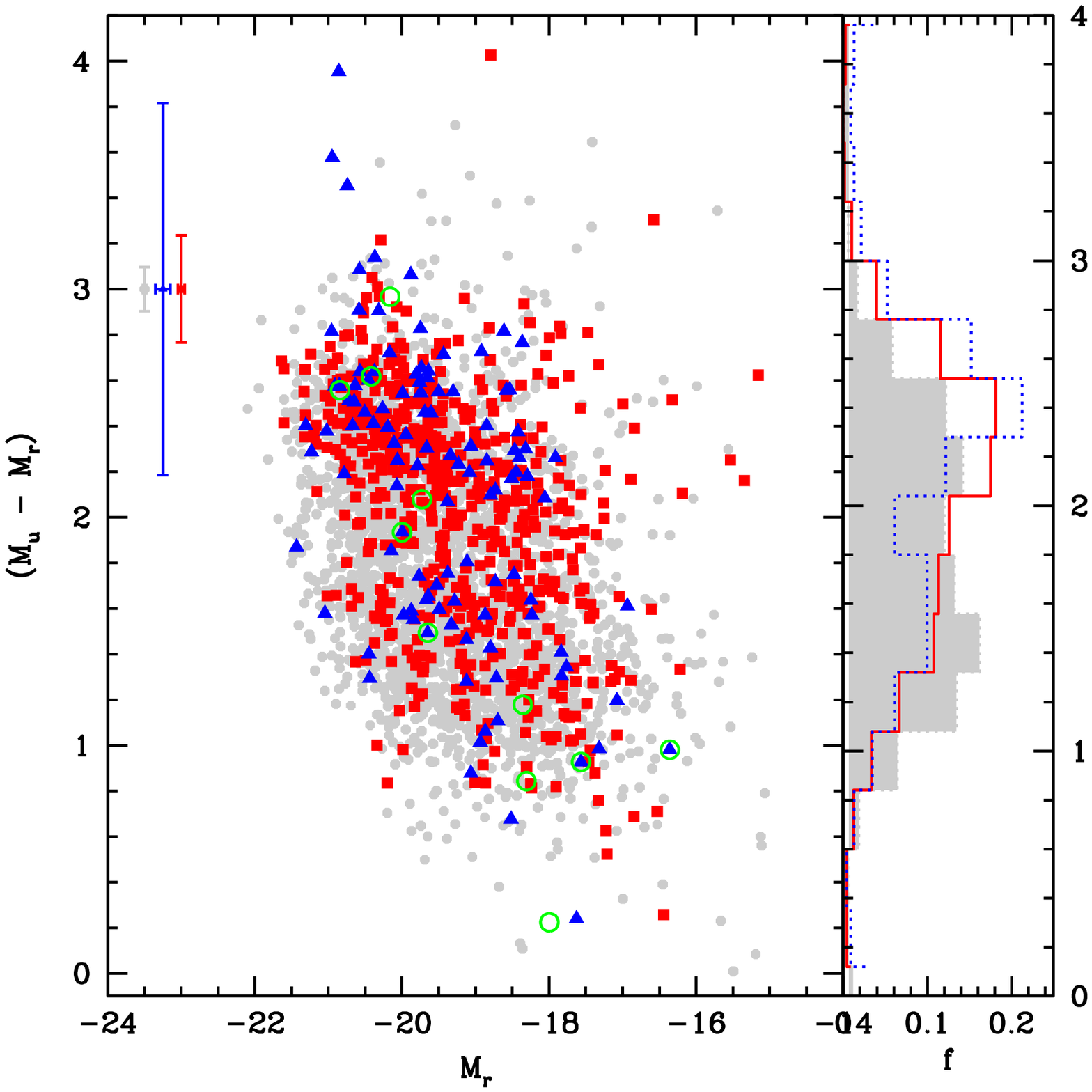}~\hfill
  \includegraphics[width=.50\textwidth]{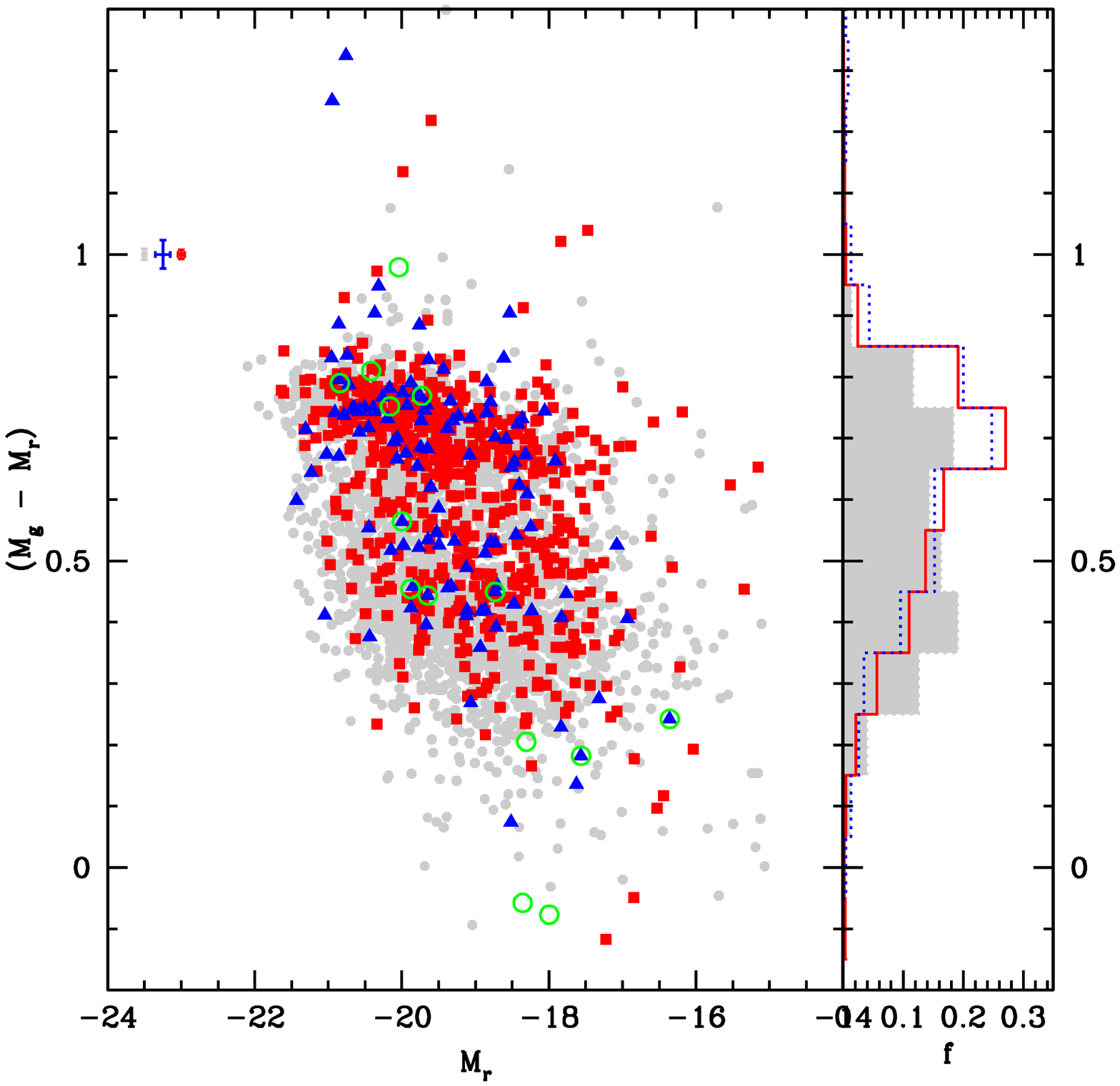}
  \caption{{Left:} Colour magnitude diagram for satellite galaxies in systems without interaction (red squares) and systems with interaction  between satellites or with main galaxy (blue  triangles) and control sample (grey dots), and $Mu-Mr$ normalised distribution. Green open circles represents satellite galaxies in systems with double interactions.  Points in the upper left corner represent data errors, for each subsample.  \protect\\{ Right:} Colour magnitude diagram  and $Mg-Mr$ normalised distribution.}
  \label{colM_s}
\end{figure*}

With the main goal of characterising the colours of the satellite galaxies belonging to our sample, in Fig \ref {colM_s} can be seen the colour-magnitude diagrams of the satellite galaxies, with and without interactions with their respective central galaxies, including the comparison sample. In it, it is observed how the galaxies belonging to systems are found to a greater extent populating the "red sequence" and "green valley", unlike the control galaxies that have a more spread distribution. In addition, the distribution of the $Mu-Mr$ and $Mg-Mr$ colour indexes, which accounts for this behaviour, is presented in the same figure.

\begin{figure}
 \centering
 \includegraphics[width=.4\textwidth]{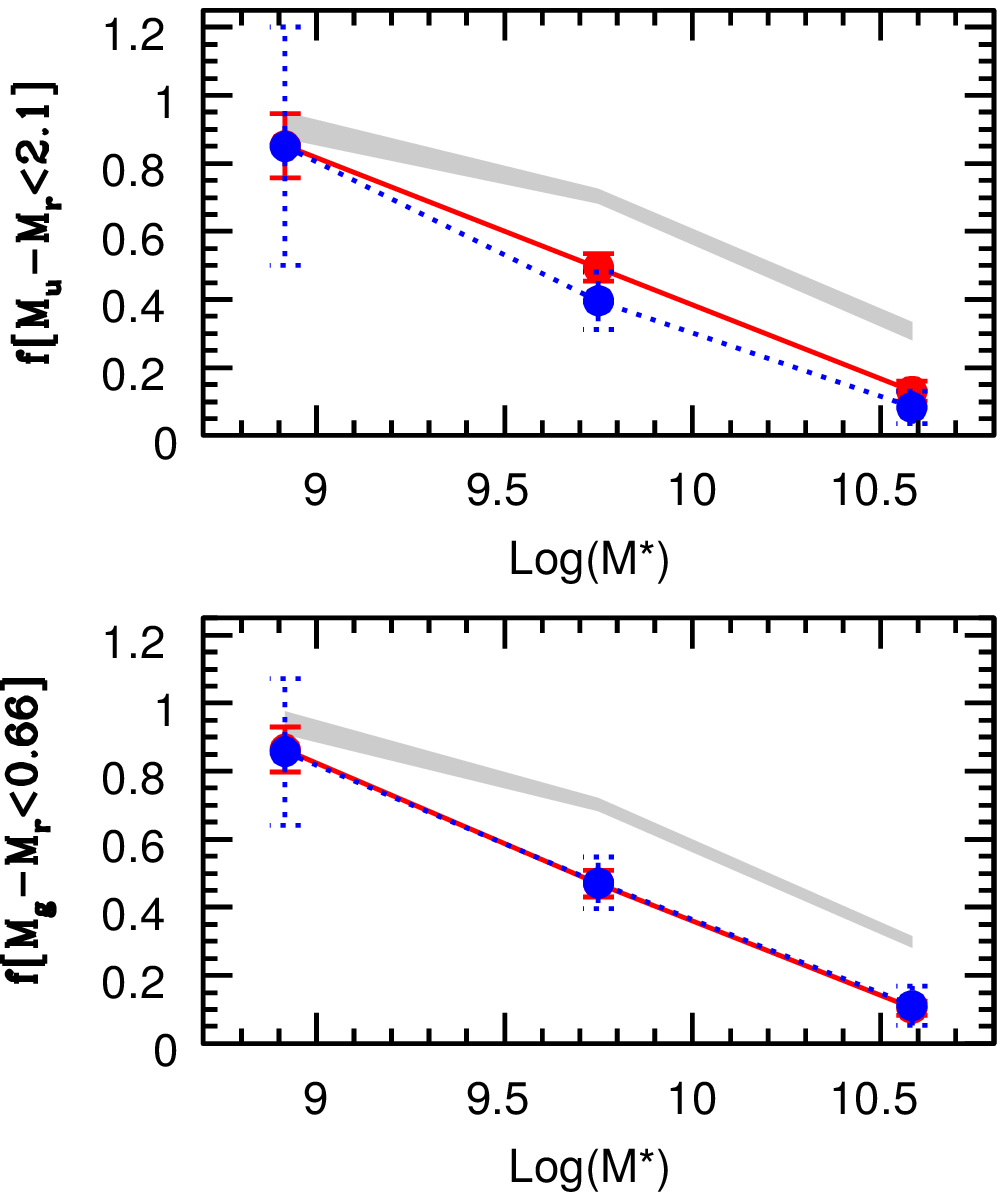}
\caption{Fraction of galaxies with $Mu-Mr $ and $Mg- Mr$  lower than the median of the sample for satellite galaxies in systems without interaction (red solid) and systems with interaction between satellites or with main galaxy (blue dotted) and control sample (shaded).}
  \label{frac_ur2}
\end{figure}

The fraction of objects with bluer colours than the median of the sample was computed for satellite galaxies. Fig \ref{frac_ur2} display this fractions as a function of $Log(M*)$, showing that satellite galaxies have a lower fraction of blue colours with respect to the comparison sample.

From these analysis we argue for an early enhancement of the star formation activity in the satellites due to the strong effect of the central galaxy, producing a rapid consumption of the gas and therefore at the present they are  redder, with older stellar population and lower SFR.
Hence, the gravitational/tidal interactions developing between satellites and central during the satellites orbits are the physical mechanism responsible for removing gas from the satellites,
causing a fast quenching of their stellar populations. This fact had already been noticed by \cite{Gutierrez2006} and \cite{Knobel2013}.

\subsubsection{Star formation and stellar populations}
\begin{figure}
\centering
\includegraphics[width=.5\textwidth]{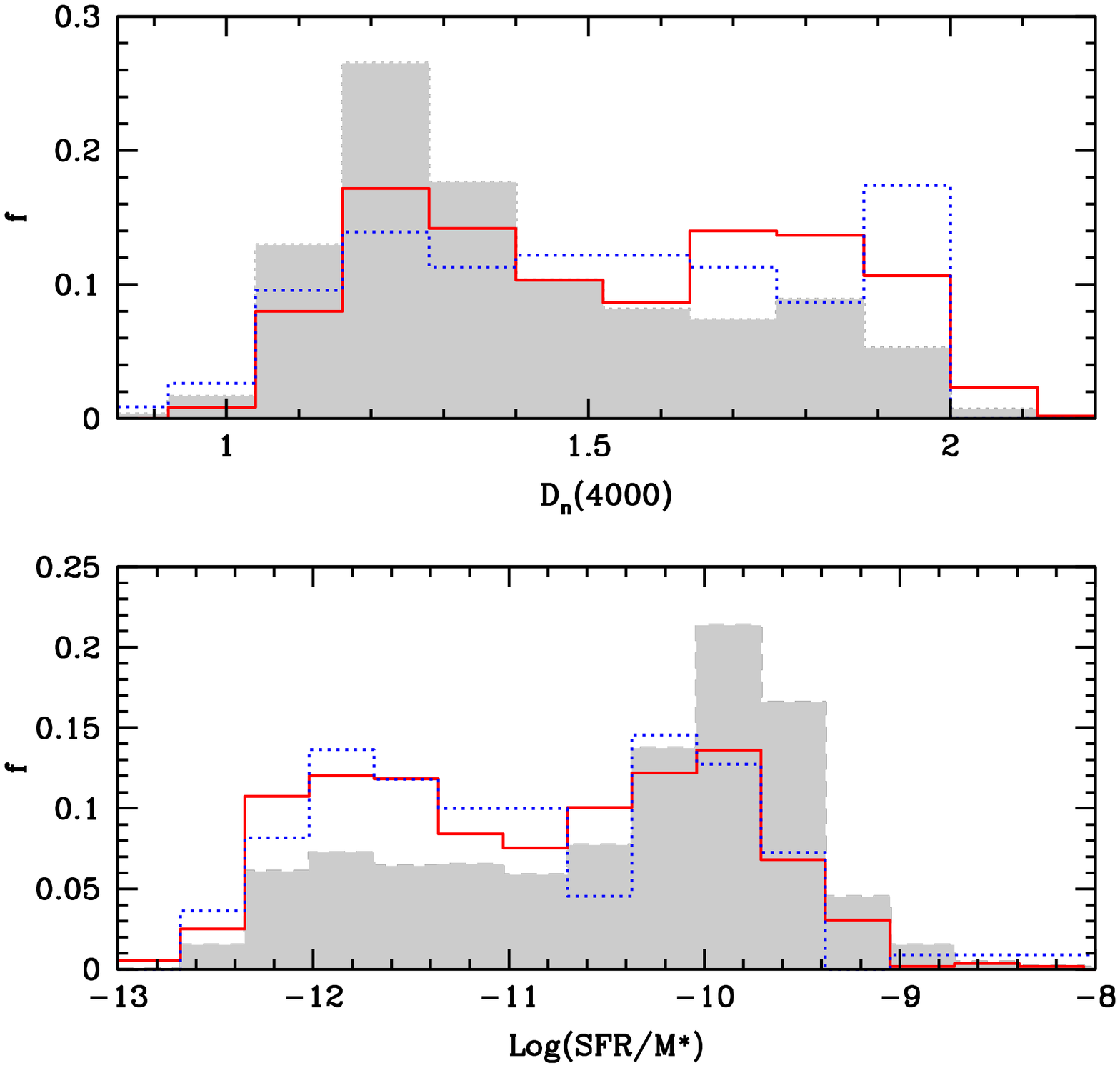}
\caption{Distribution of $D_n(4000) $  spectral index and $Log(SFR/M_*)$ for satellite galaxies in systems without interaction (red solid) and systems with interaction between satellites or with main galaxy (blue dotted) and control sample (shaded).}
\label{dn_s}
\end{figure}

Fig \ref{dn_s} shows the normalised distributions of the spectral index $ D_n (4000) $ and the specific star formation rate $ Log (SFR / M _ *) $ for each type of system. The control sample defined in Section 3.2 is also included. A bimodal distribution can be seen for the galaxies belonging to the systems under study. An important presence of galaxies with older stellar populations and low stellar formation, regardless of the type of interaction, is also observed. However, unlike the central galaxies, a counterpart with active star formation and younger populations is present. Both samples present similar behaviours, differentiating themselves from their control sample, composed of a younger population, with clear efficient star formation.

\begin{figure}
  \centering
 \includegraphics[width=.4\textwidth]{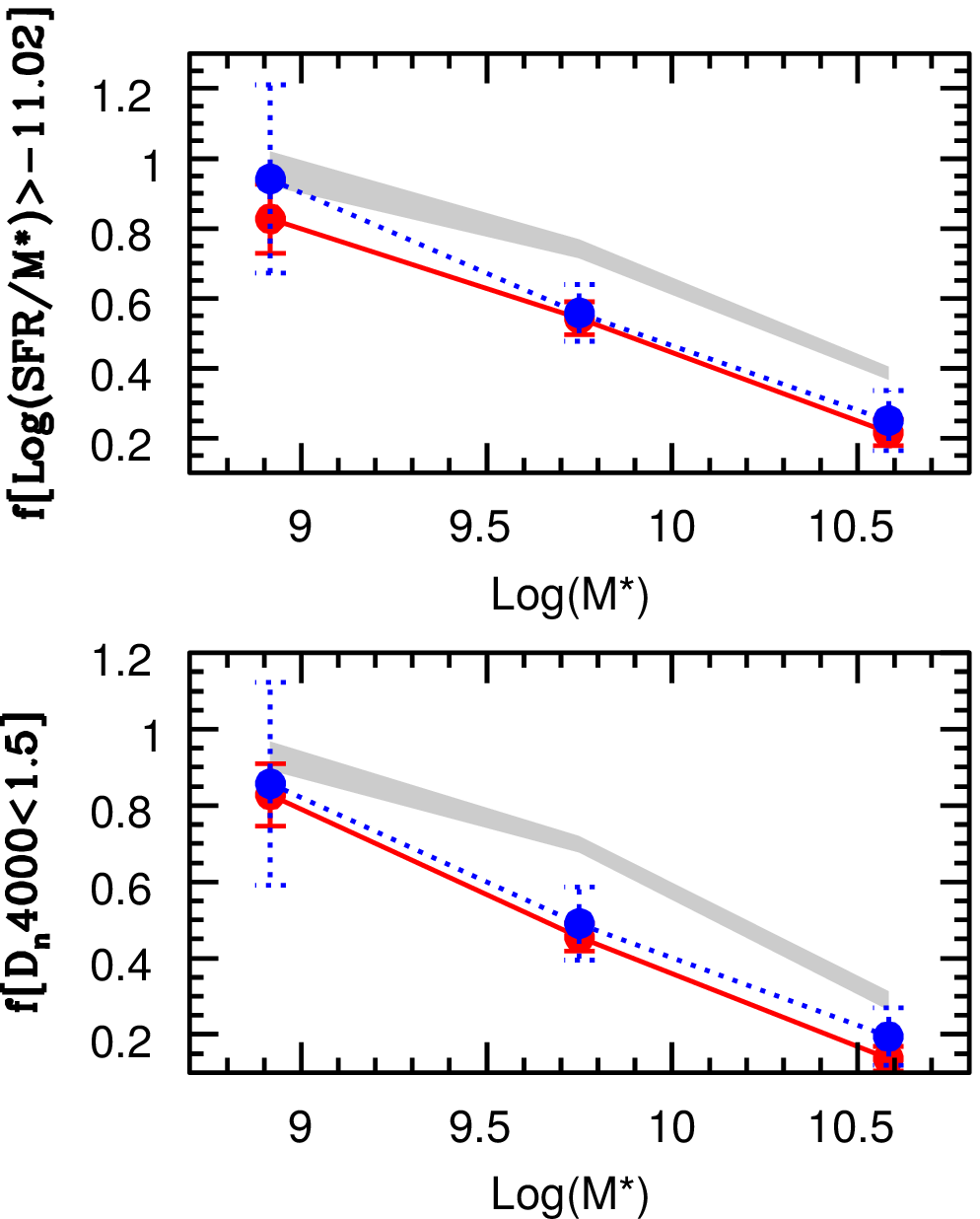}
  \caption{Fraction of galaxies with $Log(SFR/M_*)$ higher and $D_n(4000) $ lower than the median of the sample for central galaxies in systems without interaction (red solid) and systems with interaction between satellites or with main galaxy (blue dotted) and control sample (shaded)}
 \label{frac2}
\end{figure}

This behaviour is evidenced in Fig \ref {frac2}, where fractions of young galaxies with signatures of star formation are calculated. These fractions have been determined according to the median value of the total sample. These values are below the values found for the central galaxies, revealing more differences between the samples, and hence the importance of studying them independently. In this case, it is observed how these amounts increase at a lower stellar mass. However, there seem to be no differences between the samples, because regard they interact or not, both are always kept below the control sample. In view of these results, we can confirm a trend that has already been observed in previous section and by different authors  \citep[e.g.][]{vandenBosch2008, Wetzel2012, Tinker2010, Hirschmann2014, Oman2016}. Satellite galaxies are quenched with respect to the field as can be deduced from its comparison with the control sample, and strangulation would be a possible mechanism for this to occur. 

\subsection{Systems with double interactions}

\begin{figure*}
	\centering
	\includegraphics[width=0.7\textwidth]{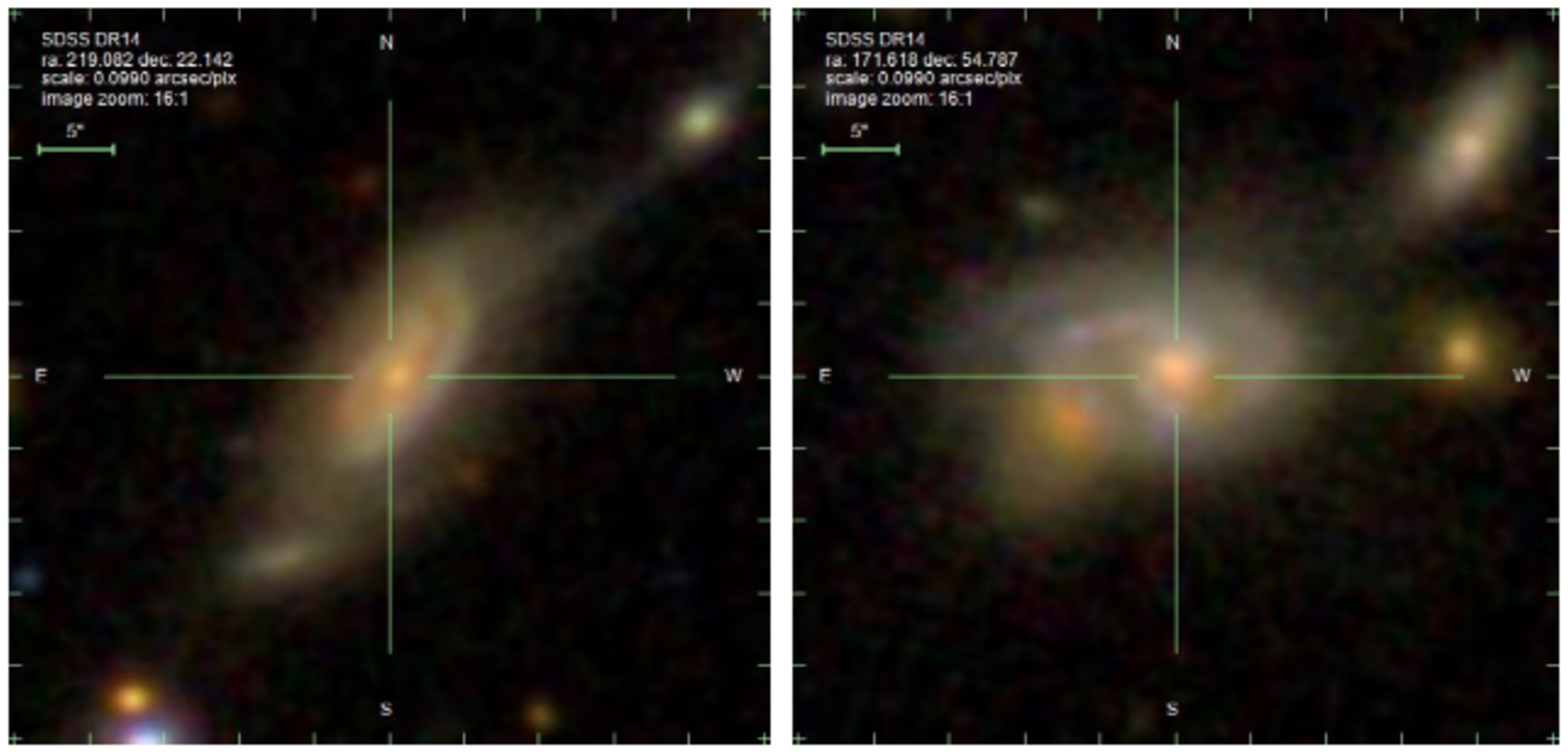}
	\caption{Examples of galaxy systems images with double interactions}
	\label{fig:example1}
\end{figure*}

Among the systems with evident interactions, there is the particular case of those in which the two satellites are in interaction with their central galaxy.  In our sample we found seven systems that present this feature. Some examples can be seen in the Fig. \ref{fig:example1}.
We think there must have special consideration with these, since the processes are likely to be more efficient, and that can be reflected in their properties. For instance, we have carried out the statistics of the indicators of star formation and age of stellar populations, finding values in the median of $Log(SFR/M_*)$ and $D_n(4000) $ of -10.19$\pm$0.26 and 1.23$\pm$0.06 respectively for satellite galaxies. In a similar way, we found median values of $Log(SFR/M_*)$ and $D_n(4000) $ of -9.80$\pm$0.14 and 1.31$\pm$0.07 severally for central galaxies experiencing double interactions.
This finding indicates that the double interaction systems are composed by galaxies, two satellites and one main object, with efficient star formation activity and young stellar populations.

The position of these objects in the colour magnitude diagrams has been highlighted (Figs. \ref{colM} and \ref{colM_s}), and they are represented by the green open circles in it. In both cases it can be seen how the galaxies involved in double interactions trace the relationship in a clearer and more linear way than the other systems. 

\subsection{Dependence on projected distance}

As shown in previous sections, we find that interactions can affect diverse galaxy properties by inducing different process. These properties may change in different ways depending on whether the galaxies are centrals or satellites. Besides, they may also correlate with the relative mass of the interacting galaxies. An important issue to explore is the dependence of the effects on projected distance between the interacting galaxies. To further explore this fact, in this section we study the fractions of younger, star-forming and bluer galaxies, according to the median of each sample. This study considers  the total sample, and a subsample of galaxies with relative projected distance $ r_p <85 \ kpc $, corresponding to the median value of the total sample.

In Table \ref{t2} the percentages of galaxies with values of $ log (SFR / M _ *) $, $ D_n (4000) $ and $ (M_u-M_r) $  below the median of their corresponding sample are computed, for the total sample and for the subsample composed by systems at closer projected separations ($ r_p <85 \ kpc $). Following the development of this work, the values are presented separately for the samples of central galaxies and their satellites.
\\

In this table the different behaviour of central galaxies and satellites in response to interactions can be clearly seen. In the case of centrals, there is always a higher percentage of those systems that have interactions, and this is even more evident for the subsample with smaller values of $ r_p $, on the other hand in systems without interactions signs, these values remain nearly constant. This is different in the case of satellites, although there is a difference in systems with interactions within the sample with lower $ r_p $, this difference is not as significant as in the previous case. Systems without interactions maintain a similar proportion according to  projected distance. By comparison with the total sample, it can be seen that these fractions are not significantly affected by the interactions.\\

\begin{table*} 
\center
\caption{Percentages of galaxies with values of $log(SFR/M_*)$,  $D_n(4000)$ and  $(M_u-M_r)$  smaller than the median of the sample for central and satellite galaxies. }
\begin{tabular}{c c c c c}
\hline
& \multicolumn{2}{c}{Systems with $r_p< 85 \kpc$} &   \multicolumn{2}{c}{Systems with $r_p< 150 \kpc$}   \\
\hline
  Ranges & Interacting $\%$ &  Non-interacting $\%$ & Interacting $\%$ &  Non-interacting $\%$    \\       
\hline
\hline
 Central galaxies & &  & &\\ 
\hline
 $log(SFR/M_*)>-12.25$ & 75.76 $\pm$ 0.87  &  48.07 $\pm$ 0.69  & 66.09 $\pm$ 0.81  &  46.18 $\pm$ 0.68  \\
\hline
  $D_n(4000) < 1.92$          &  69.70 $\pm$ 0.83     &  44.78  $\pm$ 0.67 &  64.35 $\pm$  0.80   &   44.67 $\pm$ 0.67   \\
\hline
  $(M_u-M_r)<2.45$           & 62.12 $\pm$ 0.78  &  47.76 $\pm$ 0.69  &  53.91 $\pm$ 0.73   &   45.50$\pm$ 0.67    \\
\hline 
 \hline
Satellite  galaxies & &  & &\\ 
\hline
 $log(SFR/M_*)>-11.02$ & 60.66 $\pm$ 0.78  &  51.39 $\pm$ 0.72   &  52.72 $\pm$ 0.73   &  53.76 $\pm$ 0.73  \\
\hline
  $D_n(4000) < 1.50$          &  56.06 $\pm$ 0.75     &   45.89 $\pm$ 0.67 &  48.69 $\pm$ 0.69     &    48.83$\pm$ 0.70   \\
\hline
  $(M_u-M_r)<2.10$  & 42.42 $\pm$ 0.65   &  48.51 $\pm$ 0.70   &    40.00 $\pm$ 0.63  & 50.33 $\pm$ 0.71   \\
\hline 
 
\end{tabular}
{\small  }
\label{t2}
\end{table*}

\begin{figure*}
\centering
\includegraphics[width=.45\textwidth]{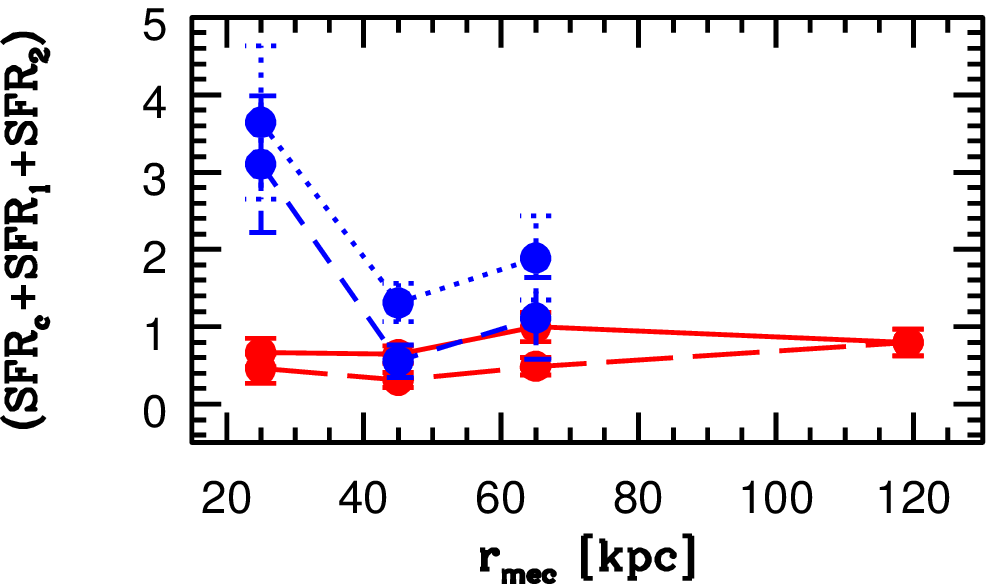}~\hfill
  \includegraphics[width=.45\textwidth]{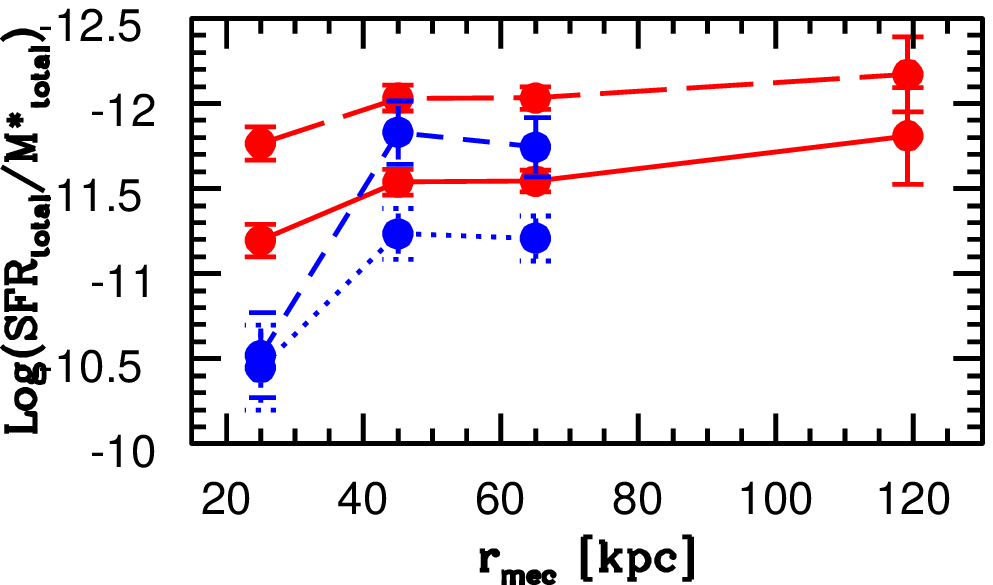}
\caption{Left:Total star formation rate ($SFR_c+SFR_1+SFR_2$) as a function of the radius of the minimal enclosing circle $(r_{mec})$ for systems of galaxies without interaction (solid) and systems with interaction between satellites or with main galaxy (dotted). Dashed lines correspond to $SFR_c$. Right: $Log(SFR_{total}/M^*_{total})$ vs $(r_{mec})$. }
 \label{sfrtot2}
\end{figure*}

On the other hand, we have estimated the minimal enclosing circle of each system, taking into account the projected distance between the three members of the system. An interesting analysis results from the study of the global properties of the systems depending on the radius of this circle. For this aim, we compute the sum of the star formation rates for the three members of a given system ($SFR_c+SFR_1+SFR_2$). Fig  \ref{sfrtot2} shows the behaviour of the total star formation rate as a function of the radius of the minimal enclosing circle $(r_{mec})$. We must emphasise that this radius expands in different ranges according to the type of interaction observed in the system, this result is expected.  Also, we plot $Log(SFR_{total}/M^*_{total})$ vs $(r_{mec})$ as a capture of the global star formation enhancement recently happened because of interactions. We can see that for systems without interactions these values remain almost constant, within the errors considered. However, in systems with interactions, the star formation increases for smaller $r_{mec}$ values. In dashed lines, the contribution of the central galaxy has been added, where it can be seen that it is the one that dominates the trend. This is likely due to the fact that since central galaxies are more massive than satellites, and in star-forming systems, SFR increases with M*.

Although not similar work have been done using systems with two satellite galaxies, these results are consistent with those found by \cite{Lambas2003,Alonso2004,Patton2013, Patton2020} where the proximity between two paired galaxies triggers their star formation activity. In this sense, it is possible speculate that the tidal effects produced by the central galaxies in both satellites could strip them of their gas reservoir, and produce efficient bursts of star formation.

\subsection{Global star formation efficiency}

\begin{figure*}
\centering
\includegraphics[width=1\textwidth]{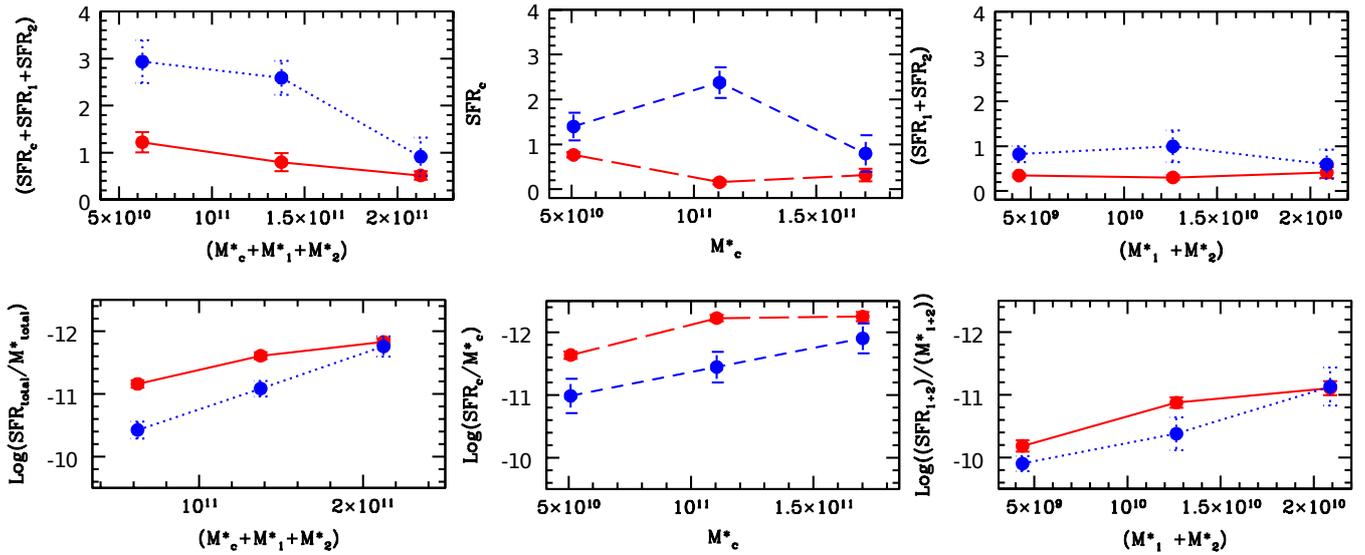}
\caption{Top: Left: Total star formation rate ($SFR_c+SFR_1+SFR_2$) as a function of total stellar mass ($M^*_c+M^*_1+M^*_2$) for systems of galaxies without interaction (solid) and systems with interaction between satellites or with main galaxy (dotted). Centre: Star formation rate $SFR_c$ as a function of stellar mass $M^*_c$ for central galaxies. Right: Total star formation rate ($SFR_1+SFR_2$) as a function of total stellar mass ($M^*_1+M^*_2$) for satellite galaxies.
Bottom: same analysis for SFR normalised to M*.} 
 \label{sfrtot}
\end{figure*}

Following the previous analysis, in this subsection we study the efficiency of interactions to trigger the formation of stars in the system considered as a whole.
To achieve this goal, we use the the sum of the star formation rates, previously calculated, and we compute the sum of the stellar masses too ($M^*_c+M^*_1+M^*_2$). Fig. \ref{sfrtot}  shows the behaviour of the total star formation rate as a function of the total stellar mass.  Also, we show $Log(SFR_{total}/M^*_{total})$ as a function of total stellar mass, analogous to the previous section, as an indicator of recent star formation.
It is clearly seen by comparison of the samples that interacting systems show an enhanced star formation activity. It can also be appreciated that this effect is more noticeable for low stellar masses, and it decreases with negligible differences for the most massive objects. 

The contribution of the effect of each member of the system can be seen in the middle and right panels of Fig. \ref{sfrtot}, for the central galaxy and for the sum of its satellite stellar contributions respectively. It can be seen that, as in the previous section, the trends are  mainly dominated by the central galaxy. However, a significant increase in the normalised SFR of low-mass satellites can be observed.

\cite{Lambas2012} results for galaxy pairs in minor interactions are in general agreement with our findings. However, we notice a possible evolution scenario where lower mass systems, more susceptible to environment, respond with significant bursts of star formation and associated colour changes. More massive systems which have more probably experienced previous interactions, are more evolved thus resembling fossil groups.

\section{Conclusions}

Using data from SDSS-DR14 we have built a sample of central galaxies accompanied by two satellites. We apply usually adopted criteria to define minor galaxy systems consistent with our previous studies and in the literature. In addition we also impose an isolation criterion to ensure that the identified systems are not affected by larger structures. 

In order to study the presence and influence of interactions in these small galaxy systems, we undertake a visual classification procedure where we considered  cases with interactions between satellites, or between satellites and the central galaxy. Satellite and central samples were studied separately and for each of these samples we constructed control samples from the galaxy catalogue having a similar $ z $, $ M_r $ and $ C $ distributions than those under consideration.

These galaxy systems will be the basis of future observational studies. In the medium term we will conduct a study in multiple wavelengths, and expand the sample with other available catalogues. In addition, these results may be used to make predictions in future high redshift catalogues. The scientific aim of these new studies is mainly focused at shedding light to the relevance of the  different mechanisms present in galaxies in close interactions. These process are key to understand the structural evolution as well as changes in stellar populations and their impact on global astrophysical characteristics \\

The main results of our analysis can be summarised as follows:

\begin{itemize}

	\item According to the selection criteria of the sample, two populations of very different galaxies have been obtained. The first composed by the central galaxies, and the second by its satellites. Among them, there are notable differences in mass and brightness, as expected.

	\item After the visual classification, it was found that around 80\% of the systems do not show evident interactions, and that the remaining 20\% does, and this mutual interaction may be between satellites or any of them with their central galaxy.

	\item The study of the central galaxies showed that this sample is composed of more evolved galaxies, and therefore with redder colours, old populations and with little star formation. This is mainly due to the nature of the sample chosen, given the criteria for its selection, which puts certain restrictions on brightness. It is observed that these luminous galaxies tend to be rather elliptical, or spirals with prominent bulges.

	\item	An analysis of the star formation and stellar populations showed that the systems that present interactions differ from the rest, with signs of recent stellar formation and younger populations. Systems without interactions behave similarly to the control sample. All these trends are decreasing as the galaxy's mass increases, the greater the mass, the difference is not observed.

	\item The analysis of the colours showed that in general and regardless of the type of interaction, these galaxies tend to be rather red. With the control sample with some more dispersion.

	\item The counterpart referred to satellite galaxies shows that these galaxies have a bimodal behaviour, with a part with old stellar populations and poor star formation, and on the other hand an important fraction of young and formative objects is observed. These values correlate directly with the mass of each object, even the least massive ones that show these signs. This behaviour is independent of the type of interaction. The control sample is always showing younger populations and greater stellar formation than the objects that are in the systems under study.

	\item With regard to colours, it is observed that the satellite sample is generally redder, and its control sample shows more blue colours.
	
	\item We have considered particularly the  case of double interactions with the central galaxy. The members of these systems show large star formation activity and young stellar populations and trace a tight colour-magnitude relation. However, it is necessary to increase the number of studied systems to confirm this trend.

	\item In order to understand the observed trends globally, an analysis dependent on the projected distance $ r_p $ was made, considering a subsample with objects with $ r_p $ less than the median value. The results found confirm the trends already observed, also highlighting the incidence of $ r_p $ in these. The results are very different for satellite and central galaxies. Highlighting mainly the central galaxies belonging to the subsample with shorter projected distances, where the fractions change significantly according to the interaction. In the total sample, the trend is maintained although to a lesser extent. On the opposite side are the satellites, in the total sample, where there are no notable differences in the percentages. If the subsample of less than $ r_p $ is taken into account, small differences can be seen.
	
	\item We have studied the global star formation efficiency of the system and its dependence on total mass and on the radius of the minimal enclosing circle of the members $(r_{mec})$. We find a strong dependence of the total SFR on these parameters for systems undergoing interactions. \\
	
\end{itemize}
	
	For all of the above, it can be concluded that both galaxy populations studied in this work are already differentiated by nature, and that makes their properties in general very different. Now, when interactions come into play, things begin to diversify, and show once again, that the effect affects in an unique way according to what place each one occupies in the system.
	
	While the central galaxies, by nature more red and passive (due to the selection constraints), when involved in an interaction rejuvenate and begin to show signs of recent star formation and younger populations, those satellite galaxies do not show differences in this aspect. 
	This supports the idea that starbursts occur but mainly in the central galaxy.
	This is also evident when compared to the control sample, galaxies with similar redshift, luminosity and morphology, but isolated in this case, always show more stellar activity than their counterpart in our systems. \cite{LaBarbera2014} showed that the star formation history of early-type central galaxies have a significant dependence on the environment, being those belonging to groups who show an stellar formation activity  that lasts in time, driven by the constant encounters with their satellite galaxies. 
	
	Additionally, it has been observed that all these trends correlate directly with the mass and projected distance between the members involved in the interaction. 
 These results support previous findings which shows that galaxy interactions are powerful mechanisms to trigger starburst and modify different galaxy properties \citep[e.g.][]{Lambas2003, Alonso2012, Mesa2014, Knapen2015, Moreno2020}. In addition, the masses and the closeness between galaxies involved in the merger are important parameters in setting the effects of the interactions \citep{Barton2000, Ellison2008, Lambas2012}.
	In this way more closed and less massive systems show efficient starbursts reflected in young stellar population and bluer colours. On the other hand, more massive systems present truncated star formation activity indicating a more evolve state. This scenario suggests that massive systems may have experienced interactions in the past and could be a previous stage of the fossil groups.

\section*{Acknowledgements} 

We would like to thank to the anonymous referee for a detailed revision of the manuscript and for the suggestions that helped to improve this paper.
This work was partially supported by the Consejo Nacional de Investigaciones
Cient\'{\i}ficas y T\'ecnicas and the Secretar\'{\i}a de Ciencia y T\'ecnica 
de la Universidad Nacional de San Juan. V.M. also acknowledges  support  from  project  Fondecyt  No.  3190736.
 
Funding for the SDSS has been provided by the Alfred P. Sloan
Foundation, the Participating Institutions, the National Science Foundation,
the U.S. Department of Energy, the National Aeronautics and Space
Administration, the Japanese Monbukagakusho, the Max Planck Society, and the
Higher Education Funding Council for England. The SDSS Web Site is
http://www.sdss.org/.

The SDSS is managed by the Astrophysical Research Consortium for the
Participating Institutions. The Participating Institutions are the American
Museum of Natural History, Astrophysical Institute Potsdam, University of
Basel, University of Cambridge, Case Western Reserve University,
University of
Chicago, Drexel University, Fermilab, the Institute for Advanced Study, the
Japan Participation Group, Johns Hopkins University, the Joint Institute for
Nuclear Astrophysics, the Kavli Institute for Particle Astrophysics and
Cosmology, the Korean Scientist Group, the Chinese Academy of Sciences
(LAMOST), Los Alamos National Laboratory, the Max-Planck-Institute for
Astronomy (MPIA), the Max-Planck-Institute for Astrophysics (MPA), New Mexico
State University, Ohio State University, University of Pittsburgh, University
of Portsmouth, Princeton University, the United States Naval Observatory, and
the University of Washington.


\section*{Data availability}

The data underlying this article will be shared on reasonable request to the corresponding author.


\bibliographystyle{mnras}
\bibliography{biblio} 

\bsp	
\label{lastpage}
\end{document}